\documentclass[12pt,preprint]{aastex}
\begin{document}
\title{The Alignments of the Galaxy Spins with the Real-Space Tidal 
Field Reconstructed from the Two Mass Redshift Survey}
\author{Jounghun Lee}
\affil{Department of Physics and Astronomy, FPRD, Seoul National University, 
Seoul 151-747, Korea}
\email{jounghun@astro.snu.ac.kr}
\author{Pirin Erdogdu}
\affil{School of Physics and Astronomy, University of Nottingham, 
University Park, Nottingham, NG7 2RD, UK}
\email{Pirin.Erdogdu@nottingham.ac.kr}
\begin{abstract}
We report a direct observational evidence for the existence of the 
galaxy spin alignments with the real space tidal field. We calculate 
the real space tidal field from the real space density field reconstructed 
recently from the Two Mass Redshift Survey (2MRS) by Erdogdu et al. in 2006. 
Using a total of 12122 nearby spiral galaxies from the Tully Galaxy Catalog, 
we calculate the orientations of their spin axes relative to the 2MRS 
tidal field. We find a clear signal of the intrinsic correlations between 
the galaxy spins and the intermediate principal axes of the tidal shears. 
The null hypothesis of no correlation is rejected at $99.99\%$ confidence 
level. We also investigate the dependence of the intrinsic correlations 
on the galaxy morphological type and the environment. It is found that (i) 
the intrinsic correlation depends weakly on the morphological type of the 
spiral galaxies but tends to decrease slightly as the type increases; 
(ii) it is stronger in the high-density regions than in the low-density 
regions. The observational result is quantitatively consistent with analytic 
prediction based on the tidal torque theory. It is concluded that the 
galaxy spin orientations may provide in principle a new complimentary probe 
of the dark matter distribution.
\end{abstract} 

\section{INTRODUCTION}

The seeds of the galaxies observed in the present universe are the tiny-
amplitude inhomogeneities of the primordial density field which are 
presumably generated by the quantum fluctuations during the inflation 
\citep{gut-pi82}. The amplitudes of the density inhomogeneities grew gradually 
via gravity after recombination to form proto-galaxies in the initially 
overdense regions. Provided that the shapes of the proto-galaxies were 
not perfectly spherical, the tidal torques from the surrounding matter 
generated the spin angular momentum from the proto-galaxies at first order 
\citep{pee69,dor70,whi84}. In consequence, the proto-galaxy spin axes came 
to be correlated with the principal axes of the local tidal shear tensors 
defined as the second derivative of the gravitational potential field 
\citep{dub92,lee-pen00}. 

The tidal effect from the surrounding matter continued till the turn-around 
moment when the proto-galaxies got separated from the neighborhood and 
began to collapse. If the spin angular momentum were well conserved after 
the turn-around moment, the spin directions of the galaxies at present 
epoch would still possess the initially induced correlations with the local 
tidal shears. In the subsequent evolution, however, it is likely that the 
complicated nonlinear processes may have modified significantly the directions 
of the spin angular momentum, decreasing the degree of the spin-shear 
alignments \citep{por-etal02}. A critical issue to address is whether 
the initially induced spin-shear alignments still remain to a non-negligible 
degree or not.

Several attempts so far have been made to measure the intrinsic alignments 
of the galaxy spins numerically \citep{lee-pen00,por-etal02,pat-etal06,
ara-etal07,bru-etal07,hah-etal07}. For instance, using the simulated galactic 
halos in recent high-resolution N-body experiments, \citet{por-etal02} 
investigated the correlations between their spin axes and the local tidal 
tensors and found that the spins of the simulated galactic halos at present 
epoch completely lost their memory of the intrinsic alignments with the 
initial tidal tensors.

In contrast, observational analyses have provided indirect evidences 
for the existence of the intrinsic alignments of galaxy spins at present 
epoch \citep{fli-god86,fli-god90,nav-etal04,ary-sau05a,ary-sau05b,tru-etal06}. 
It was \citet{fli-god86,fli-god90} who first noted the anisotropic 
orientations of the spin axes with respect to the Local Supercluster plane. 
They found that the planes of the spiral galaxies tend to be perpendicular 
to the Local Supercluster plane. Recently, \citet{nav-etal04} confirmed 
this effect, showing that the spin axes of the nearby edge-on spirals are 
inclined relative to the supergalactic plane \citep[cf.][]{ary-sau05a}.  
\citet{tru-etal06} also recently reported an observational finding that 
the spin axes of the void galaxies tend to lie on the void surfaces. 
These observational evidences are consistent with the tidal-torque picture 
that the large-scale structures have deep influences on the orientations 
of the nearby galaxy spin axes. 

Yet, these previous approaches were only indirect and still inconclusive 
about the existence of the spin-shear alignments. First of all, in previous 
approaches, the orientations of the surrounding large scale structures were 
measured not in real space but in redshift space. The redshift-distortion 
effect could cause substantial uncertainty in the final results. Furthermore, 
these previous detections of the effect of the large-scale structure on 
the galaxy spins cannot be automatically translated into the detection of 
the initial tidal effect on the galaxy spins. To find a real signal of the 
spin-shear alignments, it is inevitable to measure directly the correlations 
between the galaxy spin axes and the local tidal shear tensors. 
In fact, it was \citet{lee-pen02} who attempted for the first time to 
measure directly the spin-shear alignments. But, their approach suffered 
from noisy reconstruction of the tidal field as well as
inaccurate measurement of the spin axes of the spiral galaxies.

Very recently, the real space density field has been reconstructed from 
the densest galaxy redshift survey \citep{erd-etal06}, which may allow us 
to measure directly the intrinsic alignments of the galaxy spins with 
the tidal shear field with low statistical errors. Our goal here is to 
measure the intrinsic spin-shear correlations by reconstructing the 
real space tidal field from the 2MRS density field. 

The plan of this Paper is as follows. In \S 2, we provide a concise 
overview of the linear tidal torque model. In \S 3, we present the 
observational results of the intrinsic spin-shear correlations from the 
2MRS tidal field and the Tully Galaxy Catalog, and compare the observed 
signals with theoretical predictions. In \S 4, we summarize the 
achievements of our work and draw a final conclusion.

\section{OVERVIEW OF THE ANALYTIC MODEL}

In the linear tidal torque theory the spin angular momentum of a proto-galaxy 
is determined by its geometric shape and the tidal force from the surrounding 
matter distribution \citep{pee69,dor70,whi84}:
\begin{equation}
\label{eqn:ltt}
L_{i} = \epsilon_{ijk}T_{jl}I_{lk},
\end{equation}
where $(L_{i})$ is the spin angular momentum vector of a proto-galaxy, 
$(I_{ij})$ is the inertia momentum tensor representing the geometry of 
a proto-galactic region, and $(T_{ij})$ is the initial shear tensor 
representing the tidal torques from the surrounding matter.

In the principal axis frame of $(T_{ij})$, equation (\ref{eqn:ltt}) is 
written as
\begin{equation}
\label{eqn:spa}
L_{1} \propto (\lambda_{2}-\lambda_{3})I_{23}, \quad
L_{2} \propto (\lambda_{3}-\lambda_{1})I_{31}, \quad
L_{3} \propto (\lambda_{1}-\lambda_{2})I_{12}, 
\end{equation}
where $\lambda_{1}$,$\lambda_{2}$,$\lambda_{3}$  represent the three 
eigenvalues of $(T_{ij})$ with $\lambda_{1}\ge\lambda_{2}\ge\lambda_{3}$,  
and $I_{12}$,$I_{23}$,$I_{31}$ are the off-diagonal components of 
$(I_{ij})$ expressed in the principal axis frame of $(T_{ij})$.
Since the absolute magnitude of $(\lambda_{3}-\lambda_{1})$ is the largest, 
equation (\ref{eqn:spa}) suggests that $\vert L_{2}\vert$ is the largest 
on average provided that the off-diagonal components, 
$I_{12}$,$I_{23}$,$I_{31}$, are not zero. 

Hence, it is uniquely predicted by the linear tidal torque theory that 
the spin angular momentum of a proto-galaxy is intrinsically aligned with 
the intermediate principal axis of the local tidal shear tensor. 
As mentioned in \S 1, an important question to answer is whether this 
initially induced spin-shear alignments still remain at present epoch or not.
In the frame of the linear tidal torque model, \citet{lee-pen00,lee-pen01} 
have proposed the following generalized quadratic formula to quantify the 
expected degree of the intrinsic alignments between the spin axes of the 
galaxies and the intermediate principal axis of the local tidal tensor 
at present epoch:
\begin{equation}
\label{eqn:spin}
\langle L_{i}L_{j} | \hat{\bf T}\rangle = \frac{1+c}{3}\delta_{ij} -
c\hat{T}_{ik}\hat{T}_{kj}.
\end{equation}
where ${\bf L}\equiv ({L}_{i})$ is the galaxy spin angular momentum 
vector rescaled to be dimensionless, $\hat{\bf T} \equiv (\hat{T}_{ij})$ is 
the traceless tidal tensor rescaled to have unit magnitude, and $c \in [0,1]$ 
is a correlation parameter to measure the strength of the intrinsic spin-shear 
alignments with the nonlinear modifications taken into account. For the unit 
spin $\hat{\bf L}\equiv (\hat{L}_{i})$, the correlation parameter in equation 
(\ref{eqn:spin}) is reduced by a factor of $3/5$ \citep{lee-pen01}.
If $c$ has its minimum value of zero, it corresponds to the case that 
the nonlinear effect completely broke the initial spin-shear correlations, 
so that the present galaxy spin axes have random orientations. 

Using equation (\ref{eqn:spin}), \citet{lee-etal05} have derived the following 
probability density distribution of the orientations of the galaxy spin 
vectors  relative the tidal shear tensors:
\begin{eqnarray}
p(\cos\alpha,\cos\beta,\cos\theta) &=& \frac{1}{2\pi}\prod_{i=1}^{3}
\left(1+c-3c\hat{\lambda}^{2}_{i}\right)^{-\frac{1}{2}}\times \nonumber \\
&&\left(\frac{\cos^{2}\alpha}{1+c-3c\hat{\lambda}^{2}_{1}} +
\frac{\cos^{2}\beta}{1+c-3c\hat{\lambda}^{2}_{2}} +
\frac{\cos^{2}\theta}{1+c-3c\hat{\lambda}^{2}_{3}}\right)^{-\frac{3}{2}},
\label{eqn:pro}
\end{eqnarray}
where $\hat{\lambda}_{1},\hat{\lambda}_{2},\hat{\lambda}_{3}$ are the 
eigenvalues of $\hat{\bf T}$, and  $\alpha$, $\beta$ and $\theta$ represent 
the angles between $\hat{\bf L}$ and the major, intermediate, and minor 
principal axes of $\hat{\bf T}$, respectively. 

To quantify the preferential alignment of $\hat{\bf L}$ with the 
{\it intermediate} principal axis of $\hat{\bf T}$, we calculate $p(\cos\beta)$:
In equation (\ref{eqn:pro}) $\hat{\lambda}_{i}$'s ($i=1,2,3$) satisfy the 
traceless condition of $\sum_{i}\hat{\lambda}_{i}=0$ as well as 
unit-magnitude condition of $\sum_{i}\hat{\lambda}^{2}_{i}=1$. Therefore, 
they are well approximated as 
$\hat{\lambda}_{1}=-\hat{\lambda}_{3}=-1/\sqrt{2}$ and $\hat{\lambda}_{2}=0$. 
Putting these approximate values into equation (\ref{eqn:pro}), we derive 
the probability density distribution of $\cos\beta$ as 
\begin{equation}
\label{eqn:beta}
p(\cos\beta) = (1+c)\sqrt{1-\frac{c}{2}}
\left[1 + c\left(1 - \frac{3}{2}\cos^{2}\beta\right)\right]^{-3/2},
\end{equation}
where $\cos\beta$ is assumed to be in the range of $[0,1]$ since what 
is relevant to us is not the sign of the spin vector but the relative 
orientation. If $c=0$ the probability density distribution $p(\cos\beta)$ 
will be a uniform distribution of $p(\cos\beta)=1$; If $c > 0$, then 
$p(\cos\beta)$ will increase toward $\cos\beta=1$, indicating the existence 
of the preferential alignments of the galaxy spin axes with the intermediate 
principal axes of the tidal tensors.

The preferential alignment between $\hat{\bf L}$ and $\hat{\bf T}$ can 
be also quantified in terms of the azimuthal angle $\phi$ of $\hat{\bf L}$ 
in the principal axis frame of $\hat{\bf T}$. 
If $\hat{\bf L}$ is preferentially aligned with the intermediate principal 
axis of $\hat{\bf T}$ (i.e., $c \ne 0$), then the probability density 
distribution $p(\phi)$ should also deviate from the uniform distribution 
but increases toward $\phi = 90$ in unit of degree:
Using equation (\ref{eqn:pro}) with $\cos\alpha\equiv \sin\theta\cos\phi$ and 
$\cos\beta\equiv \sin\theta\sin\phi$, one can derive the azimuthal 
angle distribution $p(\phi)$ as 
\begin{equation}
\label{eqn:phi}
p(\phi) = \frac{2}{\pi}(1+c)\sqrt{1-\frac{c}{2}}
\int_{0}^{1}\left[1 + c\left(1 - \frac{3}{2}\sin^{2}\theta\sin^{2}\phi
\right)\right]^{-3/2}d\cos\theta,
\end{equation}
where $\phi$ is also assumed to be in the range of $[0,90]$ in unit of degree 
for the same reason explained in the above.

The tidal torque theory itself provides little guide in determining the true 
value of $c$ since its includes the nonlinear effects after the turn-around 
moment. Thus, the true value of $c$ has to be determined empirically from 
the observed galaxies. An optimal formula for the determination of $c$ was 
derived as \citep{lee-pen01}
\begin{equation}
\label{eqn:c}
c = \frac{10}{3} - 
10\sum_{i=1}^{3}\vert\hat{\lambda}_{i}\vert^{2}\vert\hat{L}^{\prime}_{i}
\vert^{2}
\end{equation}
where $(\hat{L}^{\prime}_{i})$ is the unit galaxy spin measured in the 
principal axis frame of the tidal shear. For a given sample of $N_{g}$ 
galaxies, the statistical errors involved in the measurement of $c$ was 
also found \footnote{In the original derivation of \citet{lee-pen01} 
the formula for the statistical error is found for the reduced correlation 
parameter $a = 5c/3$.} to be $\sigma_{c}=10/(3\sqrt{5N_{g}})$ 
\citep[see Appendix in][]{lee-pen01}.

\section{OBSERVATIONAL RESULTS}

\subsection{The Tully Catalog}

The optimal data for the measurement of galaxy spins would be a sample of 
spiral galaxies at low-redshifts ($z < 0.1$). The low-redshift condition is 
necessary since at high redshifts ($z\ge 0.1$) the weak gravitational lensing 
shear must cause extrinsic alignments of the galaxy spins 
\citep[e.g.,][]{cri-etal01}. Here, we adopt the Tully whole sky 
catalog as an optimal data, which complies $35,000$ local galaxies observed 
in the northern and southern celestial hemispheres with mean redshift 
of $\bar{z}\approx 0.4$ \citep{nil74,lau82}. 
For each galaxy, the catalog provides information on the 
supergalactic positions, equatorial declination ($DEC$), 
right ascension ($RA$), magnitude, velocity, redshift, morphological type, 
axial ratio ($b/a$) and position angle ($PA$). 
Among the $35,000$ galaxies, we restrict our 
attention only to spiral galaxies with morphological types of 0-9 as 
listed in Third References Catalog of Bright Galaxies 
\citep[RC3][]{dev-etal91}. The RC3 morphological types of 0-9 
correspond to the Hubble types of S0-Sm. A total of $12122$ spiral galaxies 
with median redshift of $\sim 0.02$ are selected from the Tully Catalog.

If a spiral galaxy were a thin circular disk, then the cosine of its 
inclination angle would be nothing but its axial ratio, so that its spin axis 
could be determined from  the given information on its axial ratio ($b/a$) 
and position angle ($PA$) \citep[e.g.,][]{pen-etal00,tru-etal06}. 
In practice, however, due to the disk's finite thickness and existence of 
bulge the axial ratio of a spiral galaxy is likely to be under-estimated or 
over-estimated depending on its type, which would in turn cause non-negligible 
systematic errors in the measurement of the galaxy's spin axes. 

Nevertheless, it is possible to minimize the systematic errors caused by 
the limited validity of the thin-disk approximation if the morphological 
type of the spiral galaxy is known: \citet{hay-gio84} provided corrections 
to the inclination angle, $i$, of a spiral galaxy by adding an intrinsic 
flatness parameter,$p$, as:
\begin{equation}
\cos^{2}i = \frac{(b/a)^{2}-p^{2}}{1-p^{2}}.
\end{equation}
According to \citet{hay-gio84}, the value of the intrinsic flatness 
parameter, $p$, varies with galaxy morphological type as
\begin{eqnarray}
p &=& 0.23, \quad {\rm S0-Sa} \nonumber \\
  &=& 0.20, \quad {\rm Sab}\nonumber \\
  &=& 0.175, \quad {\rm Sb} \nonumber \\
  &=& 0.14, \quad {\rm Sbc} \nonumber \\
  &=& 0.103, \quad {\rm Sc}\nonumber \\
  &=& 0.10, \quad {\rm Scd-Sdm} \nonumber 
\end{eqnarray}
The value of $i$ is set to $\pi/2$ if $b/a < p$.
 
Adopting the above correction given by \citet{hay-gio84}, the spin axis of 
a spiral galaxy in the local spherical polar coordinate system can be 
written as
\begin{equation}
\label{eqn:spin_local}
\hat{L}_{r} =  \cos i, \quad
\hat{L}_{\vartheta} = (1-\cos^{2}i)^{1/2}\sin PA, \quad
\hat{L}_{\varphi} = (1-\cos^{2}i)^{1/2}\cos PA
\end{equation}
where $(\hat{L}_{r},\hat{L}_{\vartheta},\hat{L}_{\varphi})$ represents 
the three components of the unit spin vector in the local spherical polar 
coordinate system. It is worth mentioning here that the spin vector 
determined from equation (\ref{eqn:spin_local}) suffers from the sign 
ambiguity in $\hat{L}_{r}$, as mentioned in \citep{pen-etal00,tru-etal06}. 
Since it is not possible to determine the sign of $\hat{L}_{r}$ from the 
given information of the Tully Catalog, we apply positive sign to all 
Tully galaxies here. We expect that this sign ambiguity will play a role 
of decreasing the strength of the spin-shear alignment signal.

Now, the equatorial Cartesian coordinates of the unit spin vector, 
$(\hat{L}_{1},\hat{L}_{2},\hat{L}_{3})$, can be 
determined by using the given information on $DEC$ and $RA$:
\begin{eqnarray}
\hat{L}_{1} &=&  \hat{L}_{r}\sin\vartheta\cos\varphi + 
\hat{L}_{\vartheta}\cos\vartheta\cos\varphi - \hat{L}_{\varphi}\sin\varphi, 
\nonumber \\
\hat{L}_{2} &=& \hat{L}_{r}\sin\vartheta\sin\phi + 
\hat{L}_{\vartheta}\cos\vartheta\sin\varphi + \hat{L}_{\varphi}\cos\varphi, 
\nonumber \\  
\hat{L}_{3} &=& \hat{L}_{r}\cos\vartheta - \hat{L}_{\vartheta}\sin\vartheta 
\end{eqnarray}
where $\vartheta=\pi/2-{\rm DEC}$ and $\varphi={\rm RA}$.

Rotating the spin axis from the equatorial to the supergalactic coordinate 
system, we finally measure the spin axis of each Tully spiral galaxy in the 
supergalactic coordinate system. We measure the spin-shear alignments in 
the supergalactic coordinate system since the 2MRS density field is 
defined in the supergalactic coordinate system.

\subsection{The 2MRS Tidal Field}

An optimal data for the measurements of the tidal shear tensors would be the 
linear tidal field calculated in real space. We use the data from the 
Two Mass Redshift Survey (2MRS) which is the densest redshift survey to 
date, mapping all of the sky in the infrared bands out to a median redshift of 
$z = 0.02$ \citep{huc-etal05}. By expanding the 2MRS data in Fourier-Bessel 
functions, the real-space density field, $\delta ({\bf x})$, was constructed 
on $64^{4}$ pixels in a regular cube of linear size $400h^{-1}$Mpc in 
supergalactic coordinate system \citep{erd-etal06}. 

To construct the tidal shears from the 2MRS density field, we first calculate 
the Fourier-transform of the density field, $\delta ({\bf k})$, using the 
Fast-Fourier-Transformation (FFT) method \citep{pre-etal92}. Since the tidal 
tensor is defined as the second derivative of the gravitational potential, 
the Fourier transform of the density field is related to the Fourier 
transform of the tidal shear field $T_{ij}({\bf k})$ as 
$T_{ij}({\bf k})= k_{i}k_{j}\delta ({\bf k})/k^{2}$. Then, 
we perform the inverse Fourier-transformation of $T_{ij}({\bf k})$ to 
construct the tidal shear field in real space, $T_{ij}({\bf x})$, on the 
same cube with $64^{3}$ pixels.

With the reconstructed 2MRS tidal shear field, we calculate the tidal tensor 
at the positions of the selected Tully galaxies by means of the Cloud-in-Cell 
(CIC) interpolation method \citep{hoc-eas88}. For a given supergalactic 
position ${\bf x}_{p}$ of each selected Tully galaxy, we first find the eight 
nearest pixels. And then we interpolate the tidal tensors at the eight 
pixels to evaluate the value of $T_{ij}({\bf x}_{p})$. 
We subtract the trace from $T_{ij}({\bf x}_{p})$ and normalize it to 
have unit magnitude. Diagonalizing the unit traceless tidal tensor 
$\hat{T}_{ij}({\bf x}_{p})$ at the position of a selected Tully galaxy, 
we find the three eigenvalues 
$\{\hat{\lambda}_{1},\hat{\lambda}_{2},\hat{\lambda}_{3}\}$ and the 
corresponding three eigenvectors as well 
$\{{\bf e}_{1},{\bf e}_{2},{\bf e}_{3}\}$. 

\subsection{The Observed Spin-Shear Alignments}

\subsubsection{\it The mean value of the correlation parameter}

Now that the spin vectors and the tidal tensors are all found at the 
supergalactic positions of the Tully spiral galaxies, we are ready to 
measure the alignments between the spin axes and the principal axes 
of the tidal tensors.

Let $\alpha$, $\beta$, and $\theta$ represent the angles between the 
spin axis of a given Tully galaxy with the major, intermediate, and 
minor principal axis of the local tidal tensor, respectively: 
$\cos\alpha\equiv \vert\hat{\bf L}\cdot{\bf e}_{1}\vert$;  
$\cos\beta\equiv \vert\hat{\bf L}\cdot{\bf e}_{2}\vert$; 
$\cos\theta\equiv \vert\hat{\bf L}\cdot{\bf e}_{3}\vert$. 
Here, the angle $\phi$ is the azimuthal angle of the spin axis in the 
$\hat{\lambda}_{1}$-$\hat{\lambda}_{2}$ plane. Note that the three 
angles $\alpha$, $\beta$ and $\theta$ are forced to be in the range 
of $[0,90]$ in unit of degree since what is relevant to us is not 
the sign of a spin vector but its spatial orientation relative to 
the principal axes of the tidal tensor.

\begin{figure}
\begin{center}
\plotone{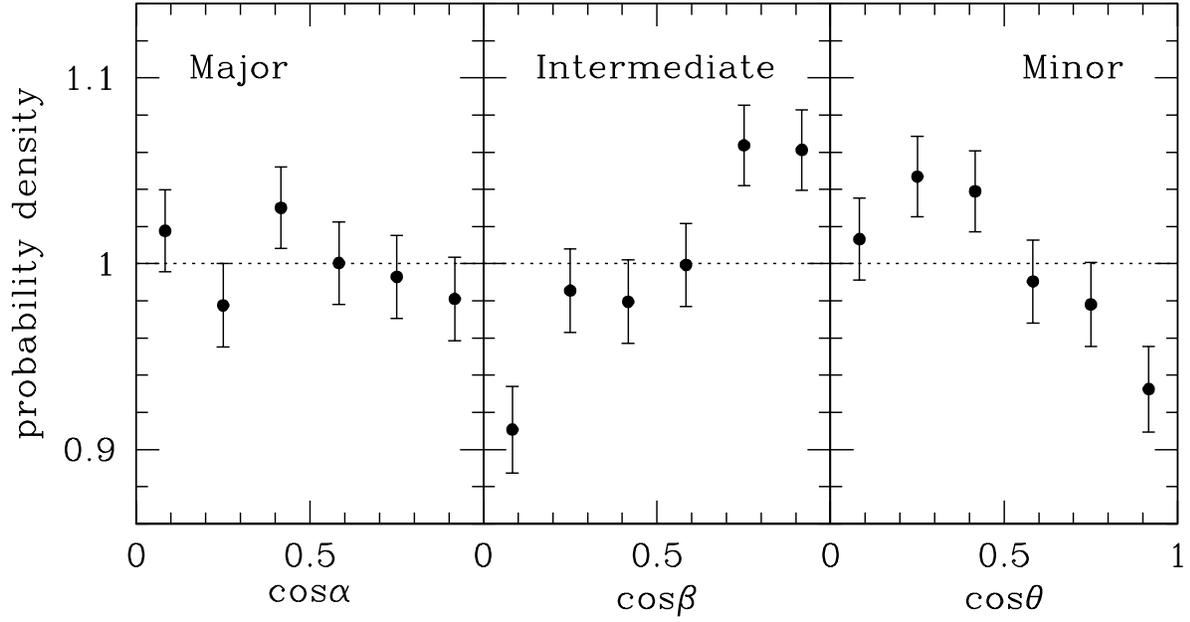}
\caption{Probability density distribution of the cosines of the 
angles between the spin axes of the Tully spiral galaxies and 
the major, intermediate, and minor principal axes of the local 
tidal tensors, in the left, middle, and right panel, respectively.
\label{fig:axi}}
\end{center}
\end{figure}

For each Tully galaxy, we calculate $\cos\alpha$, $\cos\beta$, and 
$\cos\theta$ and determine their probability density distributions.
Figure \ref{fig:axi} plots the results as solid dots with Poisson errors 
in the left, middle and right panels, respectively. 
As can be seen, the galaxy spin axes are indeed preferentially aligned with 
the {\it intermediate} principal axes of the local tidal tensors, 
which is consistent with theoretical prediction presented in \S 2.
With the help of a Kolmogorov-Smirnov test, we find that the null 
hypothesis of no spin-shear alignment is rejected at $99.99\%$ 
confidence level.

Through the similarity transformation, we express the spin vector of each 
selected galaxy in the principal axis frame of the tidal shear tensor, 
$\hat{\bf L}^{\prime}$. Then, we evaluate the correlation parameter $c$ 
by equation (\ref{eqn:c}). The mean value of $c$ is found to be 
$\bar{c}=0.084\pm 0.014$. Although the value of $c$ itself is quite low, 
it is $6\sigma_{c}$ deviation from zero, which marks a detection of the 
the intrinsic spin-shear alignments at present epoch. 

\begin{figure}
\begin{center}
\plotone{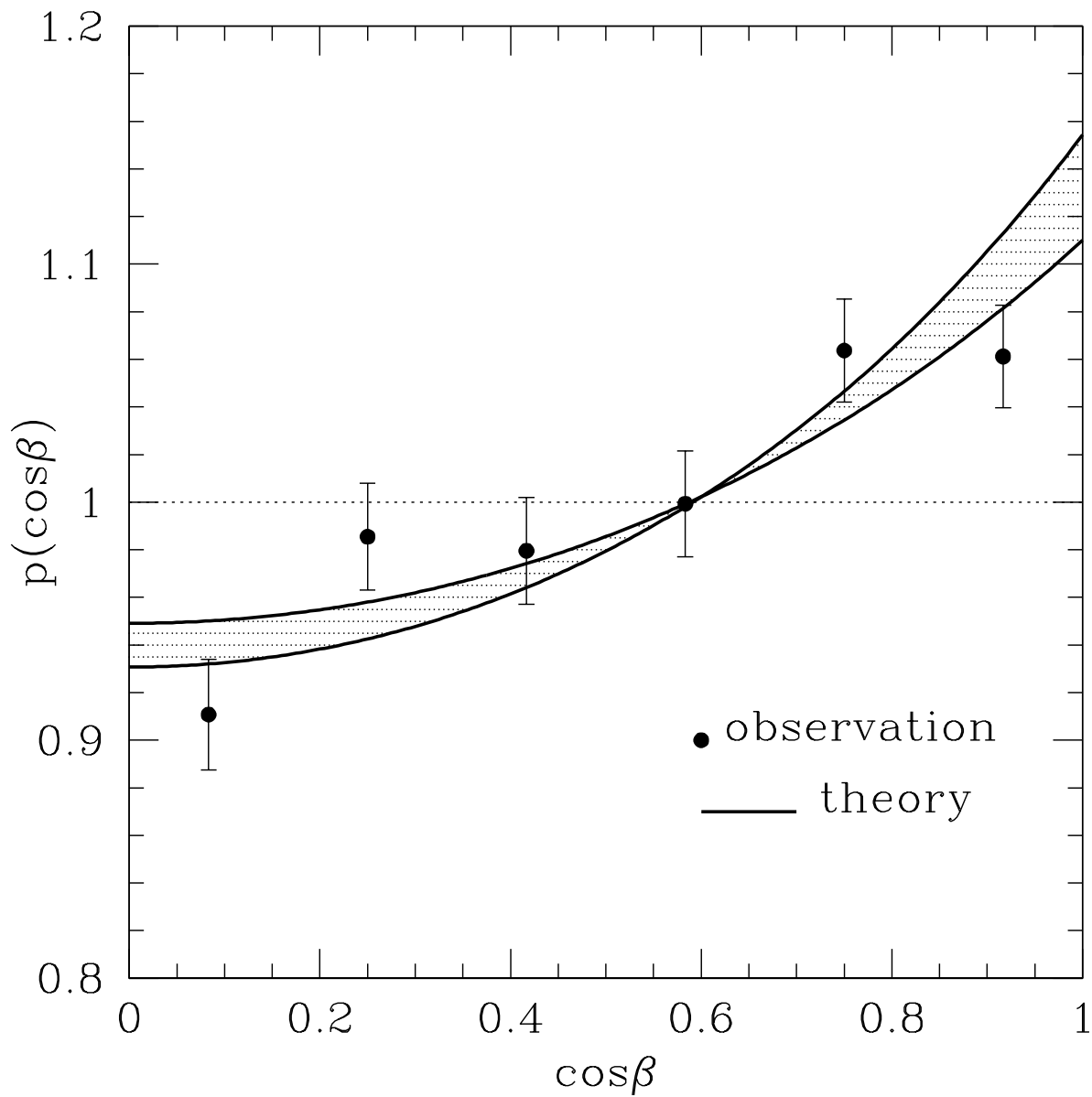}
\caption{Probability density distribution of the cosines of the angles between
the spiral galaxy's spin axes and the intermediate principal axes of the local 
tidal tensors. The solid dots with Poisson errors represent the observational 
results, the solid lines correspond to the analytic predictions, and  
the dotted line represents the case of no correlation. The shaded area 
represents $1\sigma$ of the correlation parameter.
A total of $12347$ spiral galaxies with all morphological types are used.
\label{fig:beta}}
\end{center}
\end{figure}

Figure \ref{fig:beta} compares the observational results (solid dots) 
with the analytic model (solid line). To calculate the analytic model 
we put the mean value of $\bar{c}=0.084\pm 0.014$ determined from 
the observational data into equation (\ref{eqn:beta}).  The dotted line 
corresponds to the case of no correlation. As can be seen, the analytic 
model agrees with the observational results quite well.

\begin{figure}
\begin{center}
\plotone{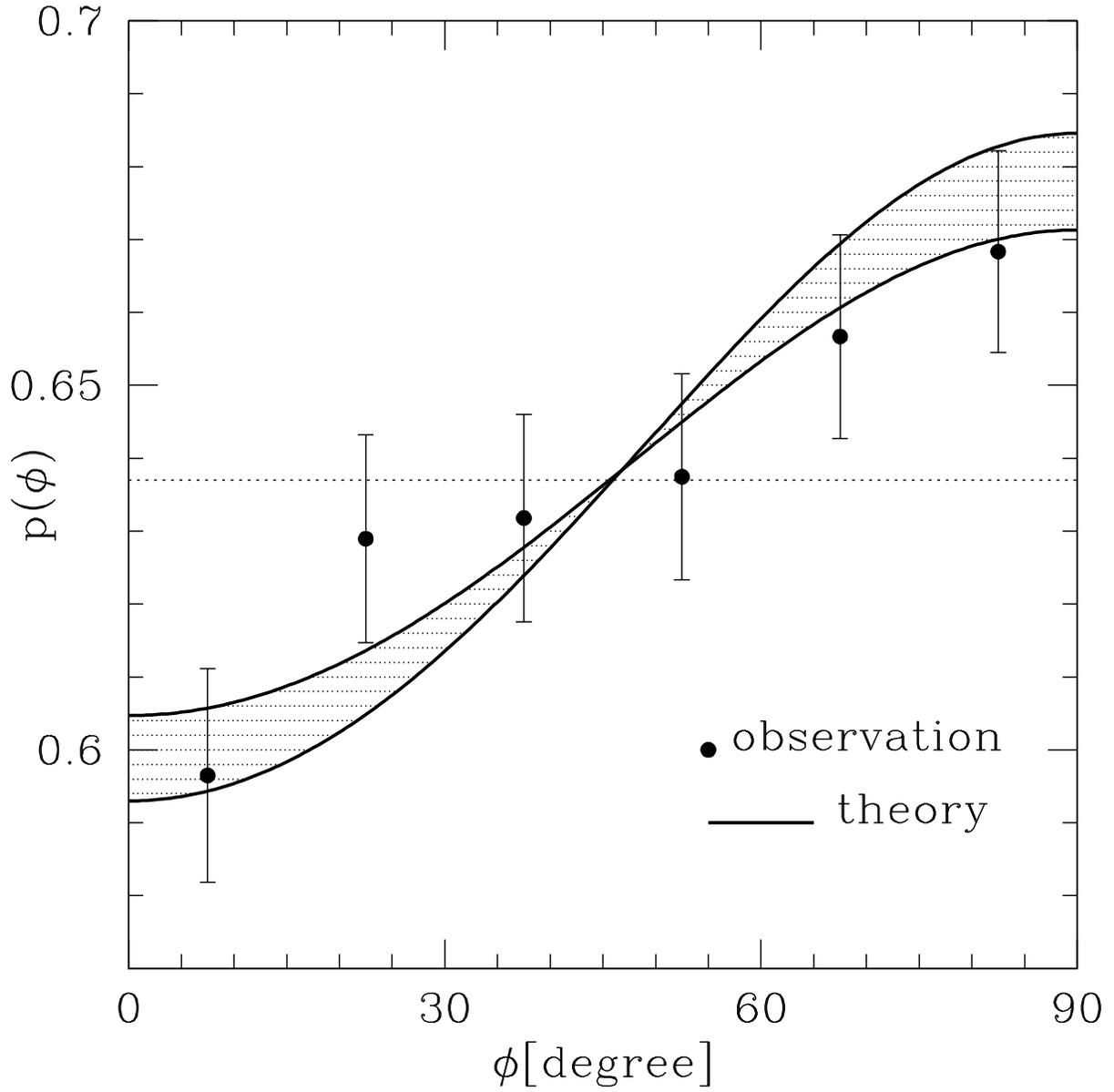}
\caption{Probability density distribution of the azimuthal angles of the 
galaxy's spin axes in the principal axis frame of the local tidal shear 
tensors. 
\label{fig:phi}}
\end{center}
\end{figure}

We also measure the probability distribution of the azimuthal angle $\phi$
of the spin axes in the tidal shear principal axis frame. 
Figure \ref{fig:phi} plots the results as solid 
dots and compares it with the analytic model as solid line (eq.\ref{eqn:phi}).
From the observational results, it is clear that $p(\phi)$ is not uniform but 
increases as the azimuthal angle increases toward $\phi=90$ in unit of degree. 
The analytic model is also in good agreement with the observational result. 
Although $p(\phi)$ is less steep than $p(\cos\beta)$, this non-uniform 
distribution of $p(\phi)$ provides an additional evidence for the preferential 
alignments between the galaxy spin axes and the intermediate principal axes 
of the tidal tensors. It reveals that the projected orientations of the 
galaxy spin axes onto the plane perpendicular to the minor principal axes 
of the local tidal field are anisotropic toward the directions of the 
intermediate principal axes of the tidal field. 

\subsubsection{\it Dependence on the morphology}

It has already been noted by previous works that the orientation 
of the galaxy spin axes may depend crucially on the galactic morphology and 
type: \citet{fli-god86,fli-god90} showed that the orientations of the galaxy 
spin axes in the Local Supercluster (LSC) tend to lie on the plane of the 
Local Supercluster, the distribution of which depends on whether the galaxies 
are seen face-on or edge-on.Recently, \citet{ary-sau05b} found that the spiral 
galaxies in the LSC are observed to have anisotropic spin orientations 
relative to the plane of LSC, while the barred spiral and irregular galaxies 
exhibit no signal of anisotropic spin orientations. Since the LSC is likely 
to have formed through the gravitational collapse along the major principal 
axes of the local tidal field, these previous results on the anisotropic 
galaxy orientations relative to the LSC provide indirect observational 
evidences for the morphological dependence of the intrinsic galaxy 
correlations with the local tidal field. 

Here, we would like to investigate more directly how the intrinsic alignment
of the galaxy spin axes with the tidal field depends on the galaxy morphology.
To investigate how the value of $c$ changes with the morphological type, 
we classify the Tully spiral galaxies into four samples with morphological 
types given as  0-1 (S0-Sa); 2-3 (Sab-Sb); 4-5 (Sbc-Sc); 6-9 (Scd-Sm). 
Then, for each sample we measure $p(\cos\beta)$, $p(\phi)$ and $c$ 
separately. 

\begin{figure}
\begin{center}
\plotone{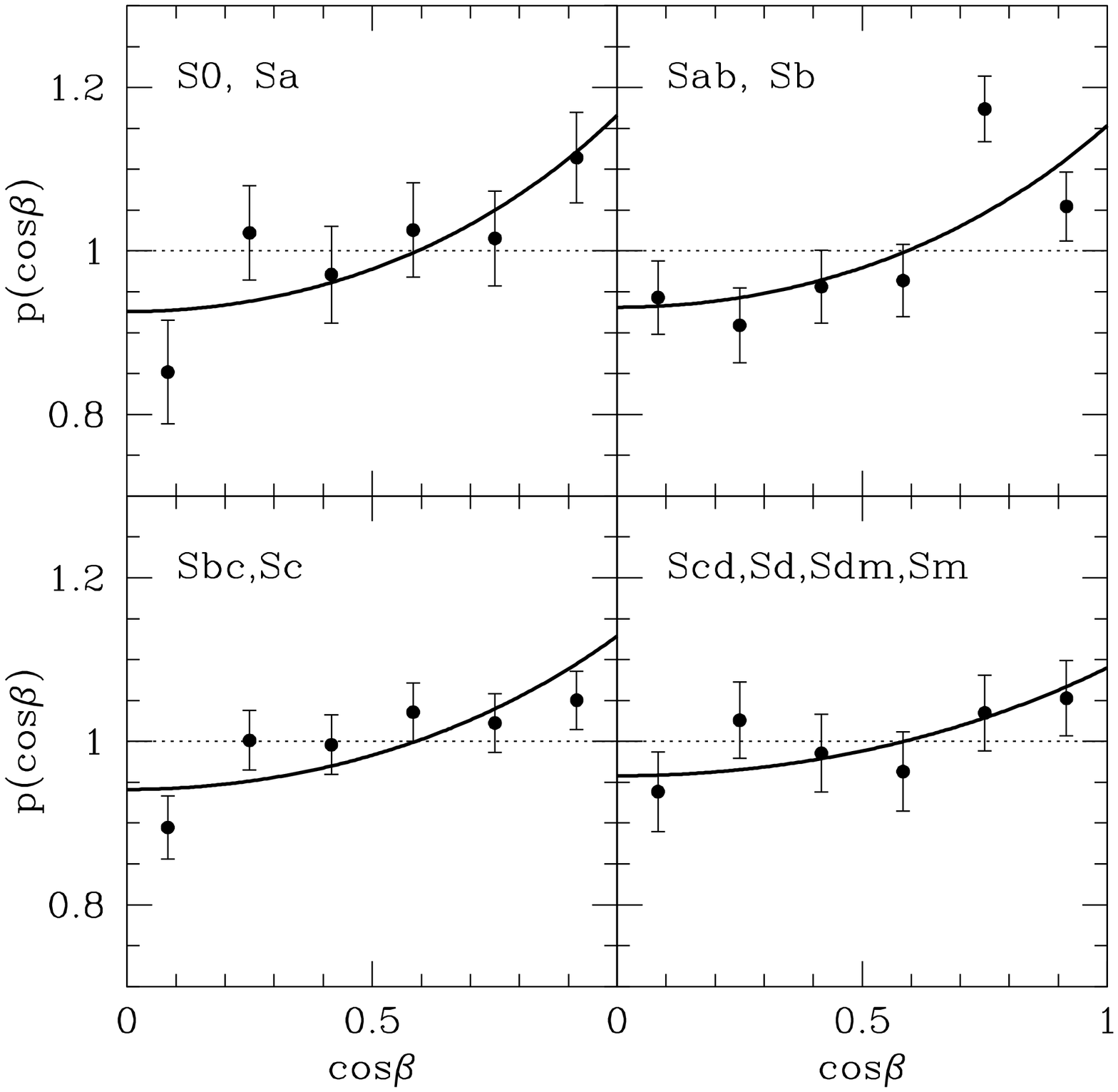}
\caption{Same as Fig. \ref{fig:beta} but for the different cases of the 
galaxy's morphological types: types of S0-Sa (top-left); 
types of Sab-Sb (top-right); types of Sbc-Sc (bottom left); types of Scd-Sm
(bottom right). 
\label{fig:typ}}
\end{center}
\end{figure}

\begin{figure}
\begin{center}
\plotone{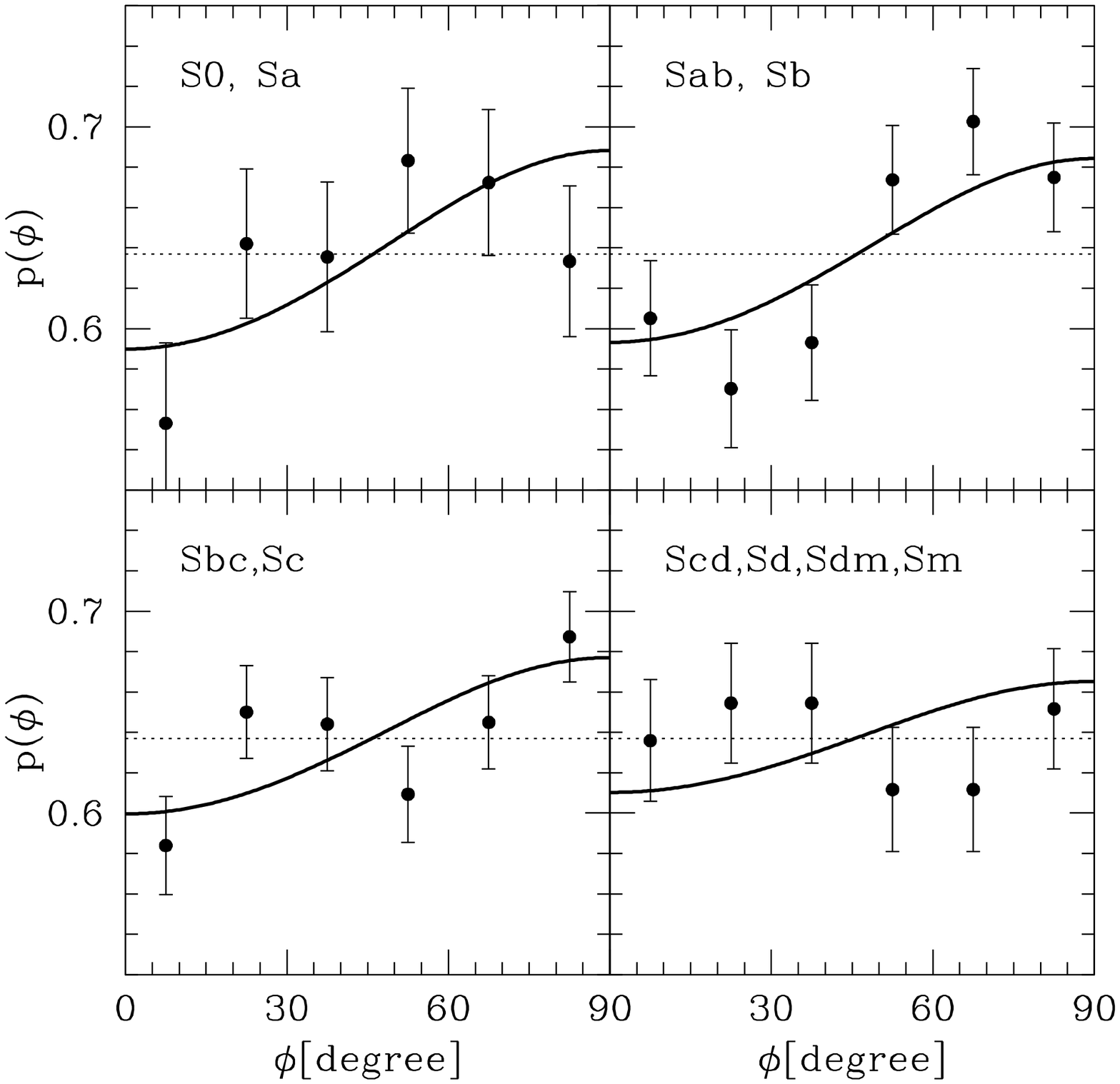}
\caption{Same as Fig. \ref{fig:phi} but for the different cases of the 
galaxy's morphological types: types of S0-Sa (top-left); 
types of Sab-Sb (top-right); types of Sbc-Sc (bottom left); types of Scd-Sm 
(bottom right). 
\label{fig:phi_typ}}
\end{center}
\end{figure}

The observational results of $p(\cos\beta)$ and $p(\phi)$ are plotted 
as solid dots in Figs. \ref{fig:typ} and \ref{fig:phi_typ}, respectively.
In each Figure, the analytic models are also plotted as solid line.
Table \ref{tab:typ} lists the number of galaxies ($N_{g}$) 
and the mean value of $c$ for the four samples. For the calculation of the 
analytic model, we use the mean value of $\bar{c}$ listed in the third column.

As can be seen, the value of $c$ does not depend strongly on the 
morphological type of the spiral galaxy. But, it tends to decrease 
slightly as the type increases. The value of the correlation parameter $c$ 
is as low as $0.05$ for the galaxy sample of types 6-9. This result suggests 
that the intrinsic spin-shear correlation is stronger for the massive galaxies.
A possible explanation for this phenomenon is that the massive galaxies are 
usually located in high-density regions and thus they experienced stronger 
tidal effect from the surrounding matter. As a matter of fact, very recently, 
\citet{lee-pen07} analyzed the numerical data from high-resolution 
N-body simulation to find that the value of $c$ increases as the mass 
of the dark halo increases. Thus, our observational result is consistent 
with numerical prediction given by \citet{lee-pen07}.

Fig. \ref{fig:phi_typ} shows that the non-uniform distribution of $p(\phi)$ 
can be seen only for the galaxy samples of types $2-5$.  However, for the 
case of $p(\phi)$ which is less steep than $p(\cos\beta)$, the large Poisson 
errors make it difficult to detect clearly the signal of the morphological 
dependence of $p(\phi)$, even though it exists, due to the small number of 
galaxies in the samples.

\subsubsection{\it Dependence on the environment}

To examine the environmental dependence of the correlation parameter $c$, 
we classify the Tully spiral galaxies into two samples belonging to the 
overdense region ($\delta<\bar{\delta}$) and to the underdense-region 
($\delta<\bar{\delta}$) where $\bar{\delta}$ is the mean density of the 
2MRS density field. Then we measure $p(\cos\beta)$, $p(\phi)$ and $c$ 
for each sample separately. 

\begin{figure}
\begin{center}
\plotone{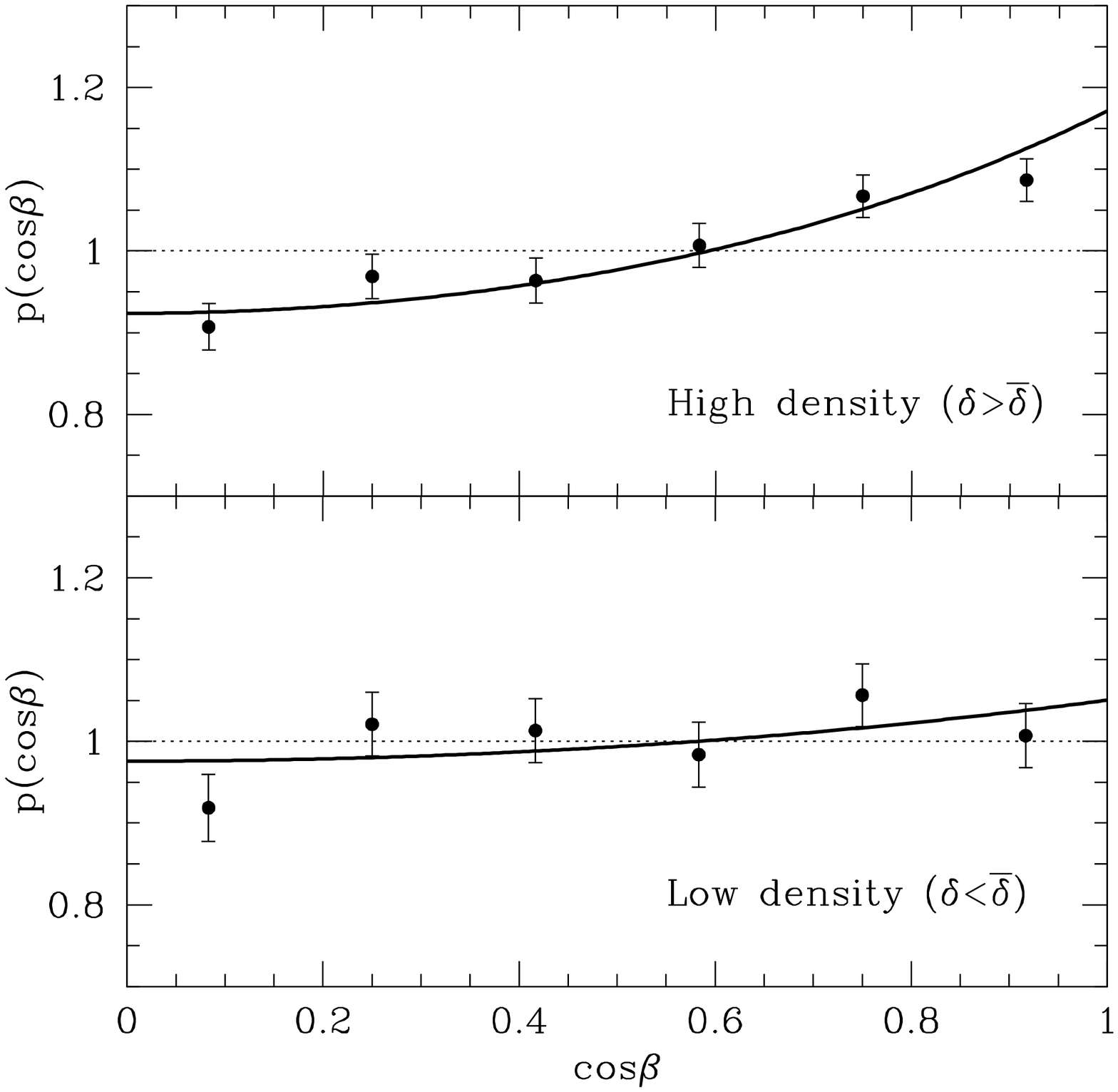}
\caption{Same as Fig.\ref{fig:beta} but for the two different cases of
the local density contrast: overdense (top) and underdense (bottom).
\label{fig:den}}
\end{center}
\end{figure}

\begin{figure}
\begin{center}
\plotone{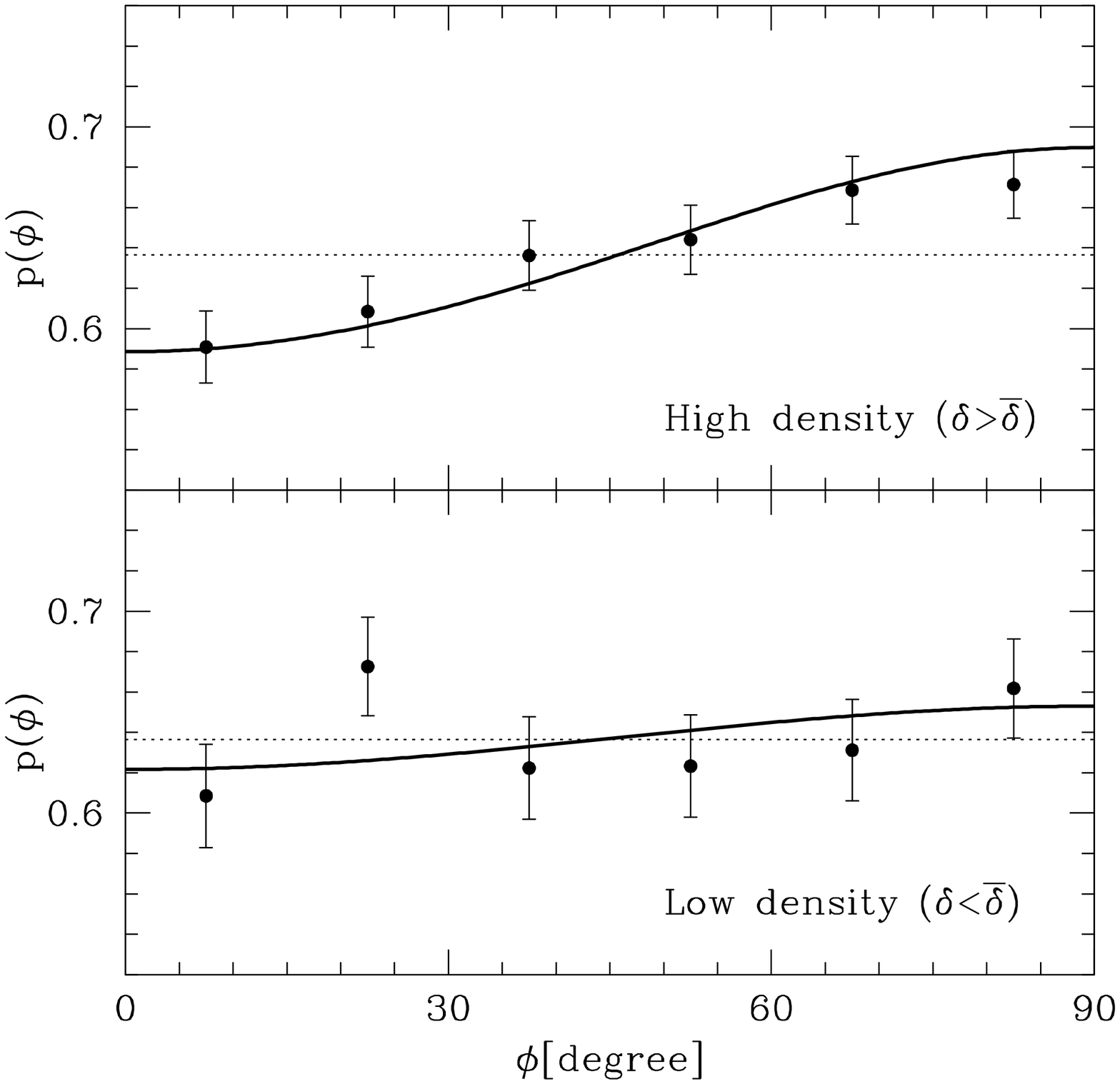}
\caption{Same as Fig.\ref{fig:phi} but for the two different cases of
the local density contrast: overdense (top) and underdense (bottom).
\label{fig:phi_den}}
\end{center}
\end{figure}

The observational results of $p(\cos\beta)$ and $p(\phi)$ are plotted 
as solid dots in Figs. \ref{fig:den} and \ref{fig:phi_den}, respectively.
In each Figure, the analytic models are also plotted as solid line.
Table \ref{tab:den} lists the number of galaxies ($N_{g}$) and the mean 
value of $c$ for the four samples. 

As can be seen, the value of $c$ is much larger in the overdense regions 
than in the underdense regions, indicating that the intrinsic spin-shear 
correlation is stronger for those galaxies which are located in the 
overdense regions. A possible explanation is that in the overdense regions 
the galaxies experience stronger tidal torques from the surrounding matter 
and thus have kept its memory of the tidal interaction better.

It is worth mentioning that in the bottom panel of Fig. \ref{fig:phi_den} 
the observational result of $p(\phi)$ has a non-negligible hump around 
$\phi = 30$ degree, which looks inconsistent with the analytic model.  
However, recall the fact that the distribution, $p(\phi)$ , corresponds to 
the alignments of the projected spin axes (as mentioned in \S 3.3.1 -2) and 
thus it is less steep than $p(\cos\beta)$. In other words, it suffers more 
severely from the small number statistics. Since the number of the galaxies, 
$N_{g}$, which belong to the low-density environment is quite small and the 
value of $c$ has a large error for this case, this hump can be interpreted 
as a statistical fluctuation within errors.

\section{SUMMARY AND CONCLUSION}

The achievements of our work are summarized as
\begin{itemize}
\item
We have measured the intrinsic alignments between the spin axes of the 
nearby spiral galaxies from the Tully Galaxy Catalog and the principal 
axes of the local tidal tensors reconstructed from the 2MRS density field.
We have detected a clear signal of the intrinsic alignments between the spin 
axes of the spiral galaxies and the intermediate principal axes of the local 
tidal tensors. The signal is statistically significant at $6\sigma$ level. 
\item
It has been found that the signal of the intrinsic spin-shear correlation 
depends weakly on the morphological type of the spiral galaxy. It is 
stronger for the early type spiral galaxies. This result is consistent 
with numerical experiment.
\item 
It has been found that the signal depends on the local density. It is 
stronger in the overdense regions, which can be understood as the 
galaxies in the overdense regions experience stronger tidal effect 
from the surrounding matter.
\item
Our results provide a compelling evidence for the tidal torque scenario 
that the galaxy spins indeed originated from the initial tidal interaction 
with the surrounding matter. 
\item
Since a significant signal of the tidal alignments of the blue 
galaxies is detected, it will be useful for the weak lensing analyses since 
it has been regarded as possible contaminants of the weak lensing signals 
\citep[e.g.,][]{hir-etal07}.
\end{itemize}

A final conclusion is that since the present galaxy spin field still keeps 
the memory of the initial tidal interaction, it is another fossil record of 
the density field at early epochs when the proto-galaxies were in expansion 
stages. Thus, the galaxy spin field can be in principle used as a new 
complimentary probe of the dark matter distribution in the universe, 
as proposed by \citet{lee-pen00,lee-pen01}

\acknowledgments

We are very grateful to an anonymous referee who helped us improve 
the original manuscript significantly. J.L. thanks B. Tully for the galaxy 
catalog and stimulating discussions. P.E. thanks the 2MRS team for 
their contributions to the reconstruction of the 2MRS density field.
J.L. acknowledges the financial support from the Korea Science and 
Engineering Foundation (KOSEF) grant funded by the Korean Government 
(MOST, NO. R01-2007-000-10246-0).

\clearpage
\begin{deluxetable}{lcc}
\tablewidth{0pt}
\setlength{\tabcolsep}{5mm}
\tablehead{Types & $N_{g}$ & $\bar{c}$}
\tablecaption{The galaxy's morphological type, the number of the 
Tully galaxies ($N_{g}$), and the mean value of the correlation parameter 
($c$).}
\startdata  
All  & $12122$ & $0.084\pm 0.014$   \\ 
S0,Sa  & $1761$ & $0.105\pm 0.036$   \\ 
Sab,Sb & $3175$ & $0.097\pm 0.026$   \\ 
Sbc,Sc & $4507$ & $0.082\pm 0.022$   \\ 
Scd,Sd,Sdm,Sm & $2679$ & $0.058\pm 0.029$  \\ 
\enddata
\label{tab:typ}
\end{deluxetable}
\clearpage
\begin{deluxetable}{lcc}
\tablewidth{0pt}
\setlength{\tabcolsep}{5mm}
\tablehead{Environment & $N_{g}$ & $\bar{c}$ }
\tablecaption{The local density, the the number of the Tully galaxies 
($N_{g}$), and the mean value of the correlation parameter ($c$).}
\startdata  
overdense ($\delta > \bar{\delta})$ & $8249$ & $0.108\pm 0.016$   \\ 
underdense ($\delta < \bar{\delta})$ & $3873$ & $0.033\pm 0.024$   \\ 
\enddata
\label{tab:den}
\end{deluxetable}


\clearpage

\end{document}